\documentclass[prl,twocolumn,showpacs]{revtex4}
\usepackage{graphicx}

\newcommand{\Dc} {D_{\rm c}}
\newcommand{\Ds} {D_{\rm s}}

\newcommand{\ie} {{i.e., }}

\newcommand{\rme} {{\rm e}}

\begin{document}

\title
{Screened hydrodynamic interaction in a narrow channel}

\author{Bianxiao Cui}
\author{Haim Diamant}
\author{Binhua Lin}
\email{b-lin@uchicago.edu}

\affiliation{The James Franck Institute, Department of Chemistry, and
CARS, The University of Chicago,
Chicago, Illinois 60637}

\date{\today}

\begin{abstract}
We study experimentally and theoretically the hydrodynamic
coupling between Brownian colloidal particles diffusing along a
linear channel. The quasi-one-dimensional confinement, unlike other
constrained geometries, leads to a sharply screened
interaction. Consequently, particles move in concert only when
their mutual distance is smaller than the channel width, and
two-body interactions remain dominant up to high particle
densities. The coupling in a cylindrical channel
is predicted to reverse sign at a certain
distance, yet this unusual effect is too small to be currently
detectable.
\end{abstract}

\pacs{83.50.Ha, 83.80.Hj, 82.70.Dd}

\maketitle

\begin{figure}[tbh]
\centerline{\resizebox{0.4\textwidth}{!}
{\includegraphics{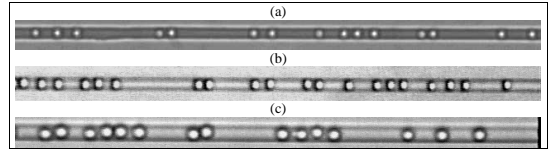}}}
\caption[]{Microscope images of silica colloidal spheres
confined in linear grooves (cf.\ Table \ref{tab_samples}):
a) sample 1, $\eta=0.17$;
b) sample 2, $\eta=0.35$;
c) sample 3, $\eta=0.36$.}
\label{fig_system}
\end{figure}

Diffusion along narrow channels is encountered in various
circumstances, such as transport in porous materials \cite{Kargerbook}
and penetration through biological ion channels \cite{ionchannel}.
Much attention has been devoted to the correlated motion in so-called
{\it single-file} systems due to the inability of particles to bypass
one another, leading to anomalous diffusion
\cite{Kargerbook,singlefile}. The anomalous regime, however, sets in
only at long enough times when particle collisions become
appreciable \cite{EPL02}.  Yet, if the channel is filled with liquid,
the motion of otherwise non-interacting Brownian particles can become
correlated at times shorter than the collision time through the flow
field that they create.  Indeed, when we observed the diffusion of
micron-size colloidal particles in a water-filled channel (Fig.\
\ref{fig_system}), the most striking feature was the occasional
concerted motion of several close-by particles in a dynamical
``train" lasting for a few seconds.  In the
current Letter we present a quantitative study of this coupling.

When a particle moves through a fluid it creates a flow that affects the
velocities of other particles in its vicinity.
Recently there has been renewed interest in such hydrodynamic interactions,
particularly in confined geometries, due to their role in the behavior
of colloidal suspensions and the development of techniques using 
digital video microscopy
\cite{sediment,Lin00,Dufresne00,Squires00,Dufresne01,Squires01}.
Colloidal particles in a finite container \cite{sediment}, near a
single wall \cite{Lin00,Dufresne00,Squires00}, and between two walls
\cite{Lin00,Dufresne01} were studied. The hydrodynamic interactions
in those geometries are always
attractive (\ie creating positive velocity correlations)
and long-ranged:
in an unbounded fluid the interaction
decays with inter-particle distance $x$ as $1/x$
\cite{HappelBrenner}; the one between particles moving near
and parallel to a single wall decays as $1/x^3$
\cite{Pozrikidis,Dufresne00}; and for particles moving between and
parallel to two walls it decays as $1/x^2$
\cite{Liron76}. More constrained geometries---perpendicular to the
walls in a two-wall configuration \cite{Liron76} and along a cylindrical
tube \cite{Liron78,Blake}---are essentially different, in that point
disturbances should create flows with an exponential spatial
decay. Here we demonstrate that confinement in a linear
channel indeed sharply screens the hydrodynamic
interaction and may even change its sign.

The experimental system consists of an aqueous suspension of silica
colloidal spheres (density 2.2 g/cm$^3$) confined in long narrow
grooves (see Fig.\ \ref{fig_system} and inset of
Fig.\ \ref{fig_collapse}).
The grooves were printed on a polydimethysiloxane substrate from
a master pattern fabricated
lithographically on a Si wafer (Stanford Nanofabrication Facility).
A drop of suspension was enclosed between the polymer mold
and a cover slip with a spacer ($\sim$100 $\mu$m),
so that the top of the groove
was open to a layer of fluid. Digital video microscopy was used
to extract time-dependent two-dimensional trajectories of the spheres
(time resolution 0.033 s).
Details of sample preparation and data analysis were described
elsewhere \cite{EPL02,JCP02}.

\begin{table}[tbh]
\begin{tabular}{|c|c|c|c|c|c|}
\hline
no. & $a$ ($\mu$m) & $w$ ($\mu$m) & $h$ ($\mu$m) & $L$ (mm) \\
\hline
1 & $0.79\pm 0.02$ & $3.0\pm 0.3$ & $3.0\pm 0.3$ & 2\\
2 & $1.85\pm 0.05$ & $5.0\pm 0.1$ & $4.0\pm 0.5$ & 10\\
3 & $2.57\pm 0.1$ & $7.0\pm 0.1$ & $4.0\pm 0.5$ & 10\\
\hline
\end{tabular}
\caption[]{Parameters of the three experimental samples.}
\label{tab_samples}
\end{table}

We studied samples of three different parameter sets, \ie different
values of sphere radius $a$, channel
width $w$, channel depth $h$, and channel length
$L$; see Table \ref{tab_samples} and Fig.\ \ref{fig_system}.
(Spheres of sample 1 were manufactured by Duke Scientific and
those of samples 2 and 3 by Bangs Laboratory.)
The ratio $a/w$ ranges between $0.26$ and $0.37$;
it is always larger than 1/4, such that spheres cannot
bypass one another.
(Higher ratios pose a difficulty in loading the spheres into the
channel.)
The linear packing fraction $\eta$ was determined as
$\eta=2Na/l$, where $l$ is the length of the channel section
in the field of view ($l=106 \mu$m for sample 1 and $220 \mu$m for
samples 2 and 3), and $N$ is the number of spheres in that section.

Studies of the equilibrium structure and single-particle dynamics of
sample 1 were previously reported \cite{JCP02,EPL02}.
These experiments show that the colloidal motion is
tightly confined to the center of the groove, with transverse
fluctuations of less than $0.2 a$.
The particle pair-potential consists of a short-ranged (screened
electrostatic) repulsion, extending to surface--surface
separations
of $\sim 0.2$ $\mu$m, followed by a weak attractive well of
$\sim -0.3 k_{\rm B}T$
($k_{\rm B}T$ being the thermal energy) at a separation of
$\sim 0.3$ $\mu$m.
At the range of the dynamic coupling reported here
the pair-potential is practically zero \cite{JCP02}.
We thus neglect in the current
study any transverse motion and direct interactions.

We choose to characterize the coupling by the collective and relative
diffusion of particle pairs, \ie the fluctuations of their
center of mass and mutual distance.
We define the collective diffusivity, $D^+$, and the relative one,
$D^-$, as
\begin{equation}
  D^\pm(x) = \langle[\Delta x_1(t) \pm \Delta x_2(t)]^2\rangle/4t,
\label{Dpmdef}
\end{equation}
where $\Delta x_i$ is the displacement of particle $i$ during time $t$
and $x$ is the short-time
average distance between the centers of the two spheres.
The coefficients have been
defined such that, in the absence of coupling, they both reduce to
the self-diffusivity of a single particle, $\Ds$.
In our data analysis $D^\pm(x)$ were calculated from the histogram
of short-time ($t<0.2$ s) trajectories, \ie the slope of the mean-square
displacements,
$
[1/4N(x)]\sum_{i=2}^{N(x)}[\Delta x_i(t) \pm \Delta x_{i-1}(t)]^2
$, as a function of $t$,
where $N(x)$ is the number of nearest-neighbor pairs whose mutual
distance $x$ falls in the range $(x-\delta x/2,x+\delta x/2)$. (We took
$\delta x=0.22a$ for all samples.) The tracking time $t$ must
be kept sufficiently short, so that the spheres do not cross over to the
anomalous subdiffusion regime \cite{EPL02}, and their diffusion distance
remains smaller than $\delta x$, so $x$ can be assumed constant.
We verified that, within experimental error, the measured
slopes were constant at least in the range
$t=0.1$--0.5 s.

\begin{figure}[tbh]
\centerline{\resizebox{0.47\textwidth}{!}
{\includegraphics{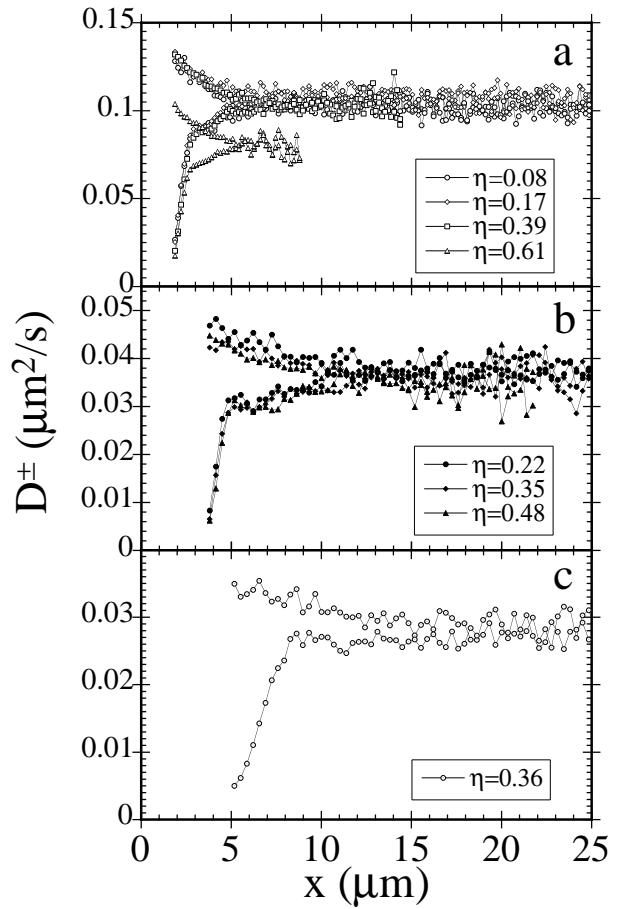}}}
\caption[]{Collective ($D^+$, upper curves) and relative ($D^-$,
lower curves) pair diffusivities as
a function of inter-particle distance $x$ for various values of
packing fraction $\eta$. Figures a, b and c
correspond to samples 1, 2 and 3, respectively.}
\label{fig_exp}
\end{figure}

Figure \ref{fig_exp} shows the measured collective and relative
diffusivities for the various samples. 
The diffusivities sharply decay to $\Ds$ for
inter-particle distances larger than about twice the channel
width. Moreover,
the coupling is practically insensitive to changes in
density up to high values of $\eta$,
implying that three-body and higher terms remain negligible.
At $\eta=0.61$ sample 1 exhibits a considerable shift of the
entire curves towards lower diffusivity, \ie a decrease in
$\Ds$ \cite{ft_Ds1}.
These observations are in essential contrast with the case of
less confined geometries, where the
hydrodynamic interaction is long-ranged with
significant many-body effects \cite{sediment}.

We now turn to a theoretical estimate of the hydrodynamic
coupling.
The Reynolds number of the system is of order $10^{-6}$.
We need, in principle, to calculate the Stokes flow due to the motion
of two particles,
subject to no-slip boundary conditions at the surfaces of the channel
and particles \cite{HappelBrenner,ft_slip}.
This is technically very hard and, furthermore, we seek a description of
more general applicability.
We resort to three simplifications:
(i) The actual
geometry of a partially open, rectangular channel is replaced by an
effective cylinder, of diameter $2R=\gamma w$, where $\gamma$ is a
geometrical prefactor of order 1, to be treated as a fitting parameter.
We expect to obtain one value of $\gamma$ for a square cross-section
(samples 1 and 2), and another for a rectangular one (sample 3).
Because of the open top of the actual channel, we also expect
the effective cylinder to be significantly wider, \ie $\gamma>1$.
(ii) The particle
motion is assumed to be restricted to the central axis of the
channel. (As has been noted above,
this is a good approximation in our case.)
(iii) The particle size is assumed to be much smaller than both the
channel width, $a\ll R$, and the inter-particle distance, $a\ll
x$. This requirement, though not
strictly fulfilled in practice, allows us to treat the effect of
one particle on the flow near the other and near the walls as if it
were exerting a point force on the fluid, \ie to consider merely the
fundamental solution (Green's function) of the Stokes flow in the
channel. (This is sometimes referred to as the {\it stokeslet
approximation} \cite{Pozrikidis}.)

The displacement fluctuations of a particle pair are used to
define a two-particle diffusion tensor,
$
  \langle\Delta x_i(t)\Delta x_j(t) \rangle = 2D_{ij}(x) t,\ i,j=1,2
$,
which can be decomposed
into a self-diffusion term and a coupling one:
$
  D_{ij} = \delta_{ij}\Ds + (1-\delta_{ij})\Dc(x)
$.
The two eigenvalues,
$
  D^\pm = \Ds \pm \Dc(x)
$,
are the collective and relative diffusivities of
the pair as defined in Eq.\ (\ref{Dpmdef}).
Usually particles entrain one another and
$\Dc>0$, \ie the hydrodynamic interaction enhances the
collective mode and suppresses the relative one.

To leading order in $a/R$ and $a/x$, the flow field induced by the
motion of one particle in the vicinity of the other is that of a
point force. This fluid velocity per unit force
is the change in the mobility of the entrained particle.
Thus $\Dc(x)\simeq k_{\rm B}T G(x)$, where
$G$ is the $xx$ component of the Stokes-flow Oseen tensor (Green's
function) in a channel. Using the
diffusivity in an unbounded fluid, $D_0=k_{\rm B}T/(6\pi\mu a)$
($\mu$ being the fluid viscosity),
we define two rescaled coupling diffusivities,
\begin{equation}
  \Delta^\pm(\xi) \equiv \frac{D^\pm(\xi) - \Ds} {(a/R) D_0},
  \ \ \ \xi\equiv x/R,
\label{rescale}
\end{equation}
which are experimentally measurable and, within our far-field
approximation, {\it parameter-free} (apart from the unknown geometrical
factor $\gamma$).

\begin{figure}[tbh]
\centerline{\resizebox{0.4\textwidth}{!}
{\includegraphics{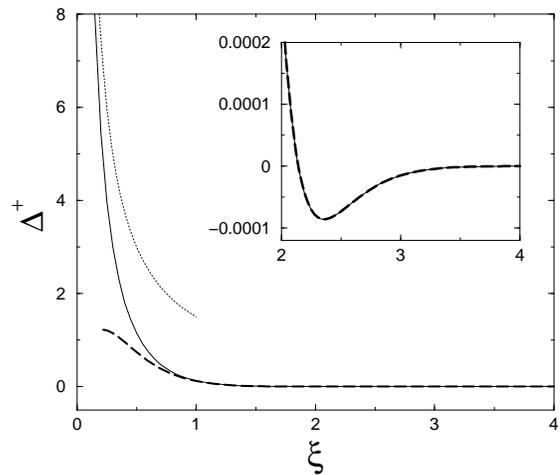}}}
\caption[]{Rescaled coupling diffusivity $\Delta^+$ ($=-\Delta^-$)
as a function
of rescaled inter-particle distance $\xi=x/R$
[Eq.\ (\ref{Delta}), solid curve].
At very small distances the curve approaches the algebraic
$\sim 1/\xi$ dependence
as in an unbounded fluid (dotted line). At large distances it
decays exponentially (dashed line). For $\xi\gtrsim 2.14$ the coupling
becomes negative (inset).}
\label{fig_Delta}
\end{figure}

\begin{figure}[tbh]
\centerline{\resizebox{0.48\textwidth}{!}
{\includegraphics{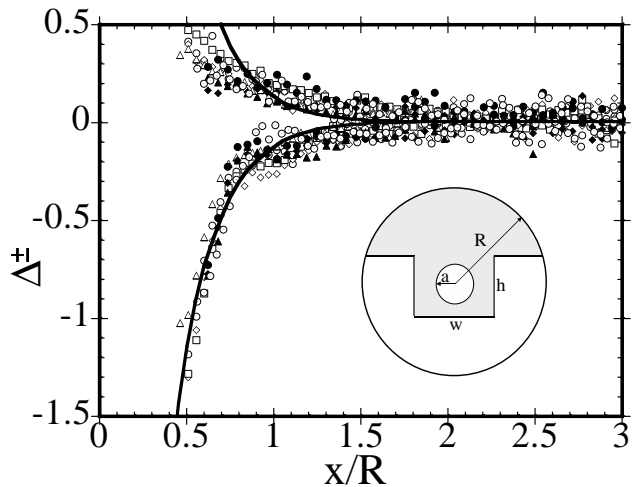}}}
\caption[]{Rescaled coupling diffusivities $\Delta^\pm$ as a function of
rescaled inter-particle distance $\xi=x/R$ for all the data of
Fig.\ \ref{fig_exp}. The solid line is the theoretical curve for a
particle in a cylindrical tube (Fig.\ \ref{fig_Delta}).
The inset shows a cross-section of the sample cell with its theoretical
effective cylinder.}
\label{fig_collapse}
\end{figure}

We now substitute the known solution for the Oseen tensor at the center
of a cylindrical tube \cite{Liron78,Blake} to obtain
\begin{eqnarray}
  && \Delta^\pm(\xi) = \pm (3/4) \sum_{n=1}^\infty
  [a_n\cos(\beta_n\xi) + b_n\sin(\beta_n\xi)]\rme^{-\alpha_n\xi}
 \nonumber\\
  &&\simeq \left\{ \begin{array}{l}
  \pm 3/(2\xi),\ \ \ \ \ \ \ \ \ \ \  \xi \ll 1 \\
  \pm (3/4)[a_1\cos(\beta_1\xi)+b_1\sin(\beta_1\xi)]\rme^{-\alpha_1\xi},
  \ \xi\gg 1
  \end{array} \right.
\label{Delta}
\end{eqnarray}
In Eq.\ (\ref{Delta}) $u_n=\alpha_n+i\beta_n$ are the complex
roots of the equation
$
  u[J_0^2(u)+J_1^2(u)] = 2J_0(u)J_1(u)
$,
and
$
  (a_n+ib_n) = 2\{ \pi [ 2J_1(u_n)Y_0(u_n) - u_n(J_0(u_n)Y_0(u_n)+J_1(u_n)
  Y_1(u_n)) ] - u_n \}/J_1^2(u_n)
$
[$J_k$ and $Y_k$ being the Bessel functions of the first and
second (Neumann) kind]. The coupling as a function of distance is depicted
in Fig.\ \ref{fig_Delta}.
For $\xi\ll 1$ the particles
are insensitive to the walls and the coupling approaches
the algebraic $\sim 1/\xi$ dependence as in an unbounded fluid
\cite{HappelBrenner}. Yet, for $\xi>1$ the confined geometry
becomes manifest; the sum is dominated by its first term and the
interaction decays exponentially with distance. (The coefficients
of this term are $a_1\simeq -0.0370$, $b_1\simeq 13.8$,
$\alpha_1\simeq 4.47$, and $\beta_1\simeq 1.47$.)

The sharp screening beyond the confinement
length is unique to the linear geometry and is caused by
the boundary conditions at the channel walls. We may
visualize the boundary conditions as replaced by an infinite
series of image forces transverse to the channel axis,
accompanying the moving particle \cite{Pozrikidis}. At distances
larger than the channel width the images cancel the effect of the
actual force. Small oscillations on top of the exponential decay
make the coupling change sign at
$\xi\simeq 2.14$ (Fig.\ \ref{fig_Delta}, inset). This is a
very peculiar effect, implying that for large distances the
particles inhibit each other's motion rather than aid it.
Physically, this arises from flow rolls that form along the channel
\cite{Blake}. Unfortunately, as seen in Fig.\ \ref{fig_Delta}, even
if the effect still exists in our partially open channel, it
is far too small to be currently detectable \cite{ft_negative}.

In Fig.\ \ref{fig_collapse} we have replotted the experimental
data of Fig.\ \ref{fig_exp}, scaled according to Eq.\
(\ref{rescale}), with $\gamma=2.44$ for the square cross-section
(samples 1 and 2) and $\gamma=2.72$ for the rectangular one
(sample 3).
The data for different channel widths, sphere
sizes, and densities collapse onto two universal curves for the
collective and relative diffusivities. The collapse confirms that the
observed coupling is well described by a two-body hydrodynamic
interaction within a first-order approximation in $a/w$.
Moreover, considering our crude approximations, the agreement
between the universal curve and Eq.\ (\ref{Delta}) is remarkable.
The stokeslet approximation thus provides a surprisingly
reasonable description
of the hydrodynamic interaction even for relatively large
particles and short inter-particle distances.
We can further use the fitted $\gamma$ to calculate the
expected self-diffusivity which, at the same level of
approximation \cite{HappelBrenner}, is given by
$
  \Ds \simeq D_0(1-2.10444 a/R),\ R=\gamma w/2
$.
This yields for samples 1, 2 and 3
$\Ds\simeq 0.16$, $0.045$ and $0.040$ $\mu$m$^2$/s, respectively.
Although these values are not far off the measured (low-density) ones---0.11,
0.036 and 0.028 $\mu$m$^2$/s---the differences,
compared to the good agreement in Fig.\ \ref{fig_collapse},
indicate that the self-diffusivity is more sensitive
than the two-particle coupling to higher orders in $a/w$ \cite{ft_Ds2}.

A small but systematic discrepancy between the
calculation and experiment is seen
in Fig.\ \ref{fig_collapse} as
the distance becomes very small.
This deviation
naturally marks the breakdown of the $x\gg a$ assumption. The observed
trend can be understood by considering the extreme case of two
particles in contact. The relative
diffusivity then vanishes, hence $\Delta^-\rightarrow -\Ds/(D_0
a/R)$, whereas the collective one becomes equal to twice
the self-diffusivity of a rigid ``train" of two touching particles,
$2D_2$.
Since, evidently, $D_2<\Ds$,
we have $\Delta^+\rightarrow (2D_2-\Ds)/(D_0
a/R) < |\Delta^-|$, in accord with the observed asymmetry.

Another important consequence of the screened coupling is that
concerted motion of neighboring particles
becomes appreciable at a sharply defined density,
comparable to $1/R$;
this dynamic clustering will be studied in more detail in a future
publication.
Our results suggest that a tube-stokeslet approximation
[\ie Eq.\ (\ref{Delta})] should give a good estimate of the
hydrodynamic interaction in {\it any} linear geometry, provided
that the particles are repelled from the walls and one has
an estimate for the geometrical factor $\gamma$.
Although we have discussed only freely diffusing particles, since
mobilities are proportional to diffusivities, the same
conclusions apply to driven motion as well \cite{Zheng},
and should also bear upon motion and patterning
in microfluidic systems \cite{microfluid}.

\begin{acknowledgments}
We thank Stuart Rice for generous laboratory support,
and
David Andelman, Cristian Huepe, Leo Kadanoff, Detlef Lohse, Tom Witten,
and Wendy Zhang for helpful comments.
The experimental work was supported by the National Science Foundation
[DMR-9870437 (POWRE) and CHE-9977841] and the NSF
MRSEC program at the University of Chicago (DMR-9808595).
HD was also supported by the NSF
(DMR-9975533 and DMR-0094569),
and the David and Lucile Packard Foundation (99-1465).
\end{acknowledgments}



\begin{thebibliography}{99}
\bibitem{ft_Ds1}
In the way $\Ds$ is measured here
it corresponds to a particle that has no neighbors on one
side but can have nearby particles on its other side. The change
in $\Ds$ at high density arises from a high probability for a third
sphere to be located less than the screening length away.
\bibitem{ft_negative}
Sign reversal of hydrodynamic interactions under
confinement is a real effect. We have recently measured negative transverse
couplings in a two-plate geometry (to be reported elsewhere).
\bibitem{ft_Ds2}
A good estimate for $\Ds$ can also be obtained
by superposing the effects of the three walls \cite{EPL02}.
\end{thebibliography}
\end{document}